\documentstyle[sprocl]{article}

\def\ra{\rightarrow}
\def\a{\alpha}
\def\b{\beta}
\def\r{\rho}
\def\s{\sigma}
\def\be{\begin{equation}}
\def\ee{\end{equation}}
\def\ba{\begin{eqnarray}}
\def\ea{\end{eqnarray}}
\newcommand{\DD}{\Delta}

\input epsf
\begin{document}

\title{Electromagnetic Form Factors and the Localization of Quark
Orbital Angular Momentum in the Proton}

\author{JOHN P. RALSTON and ROMAN V. BUNIY}

\address{
Department of Physics and Astronomy, University of Kansas,
Lawrence, KS 66045, USA}

\author{ PANKAJ JAIN}

\address{
Department of Physics, IIT Kanpur, Kanpur-208 016, India
}


\maketitle

\abstracts{ A new picture is given of generalized parton distributions
probed in experiments in which the probe scale $Q^{2}$ and the
momentum transfer $\DD^{2}$ are well separated.  Application of this
picture to the $Q^{2}$ dependence of the form factors $F_{1}, \,
F_{2}$ shows that gauge invariant quark orbital angular momentum can
be measured and indeed {\it localized} in the transverse profile of
the proton.  A previous prediction that
$\sqrt{Q^{2}}F_{2}(Q^{2})/F_{1}(Q^{2}) \sim const.  $ is generalized 
to GPD language.  This prediction appears to have been confirmed by
recent CEBAF data.}


\section{ A Physical Picture of Generalized Parton Distributions}

Deeply inelastic scattering experiments contradict the notion that the
proton's spin is the sum of the spins of the quarks.  The s-wave
non-relativistic bound state picture of the proton has then been ruled
out.  Consequently there is great interest in the orbital angular
momentum of quarks and the spin of gluons.  While the spin of gluons
is difficult to define gauge invariantly, the orbital angular momentum
of quarks is well defined in terms of generalized parton distributions
(GPD).  One of the early motivations for introducing
GPDs,~\cite{Jainral93,GPDclub} predating the current focus on virtual
Compton scattering, was the ability to express the role of orbital
angular momentum in exclusive reactions.  Rather than concentrating on
the {\it total} orbital angular momentum, obtained via a sum rule that
is probably unobservable,~\cite{JiHoodbhoy97Prl} we discuss the {\it
localization} of the orbital angular momentum region-by-region across
the transverse profile of the proton, which not only contains more
information, but is perfectly observable.  Our method extends to a
useful physical picture of the uses and purposes of GPDs.

{\it Translational Symmetry: } Consider the quark-proton scattering
amplitude $$quark(k) + proton(P) \ra quark'(k') +proton'(P+\Delta).$$
The density matrix elements describing this scattering\cite{Ralston79}
will be defined as $\Phi_{\rho \sigma}(k,k')_{P,P'}$, which in terms
of the quark fields is given by $$\Phi_{\rho \sigma}(k,k')_{P,P'}=\int
dz dz' e^{ik\cdot z-ik' \cdot z'}<P',s'| T \psi_{\rho}(z') \bar
\psi_{\sigma}(z)|P,s>.$$ By momentum conservation $k-k'=P'-P=\Delta$.
The quark fields are evaluated at spatial coordinates $z^{\mu},
z'^{\mu}$ and have Dirac indices $\rho, \sigma$.  We may rewrite the
Fourier factors via $$ e^{ik\cdot z-ik'\cdot z'}=
e^{i\frac{k+k'}{2} \cdot (z-z') +i \Delta \cdot \frac{z+z'}{2}}.$$

The proton state can be expressed as $$|P,s> = \int d Y \; exp( -iPY)
|Y,s>,$$ where $Y$ is a center of momentum ($CM$) coordinate.  The
coordinate $Y$ parameterizes collective variables of the state
without specifying more about the internal coordinates.  With a
similar step for $|P',s'>$, matrix elements depend on $e^{iP\cdot
Y-iP'\cdot Y'}= e^{i\frac{P+P'}{2}\cdot (Y-Y') -i \Delta\cdot \frac{
Y+Y'}{2} }.$ This isolates all dependence on the variable $\Delta$ in
\ba \Phi_{\rho \sigma}(k,k')_{P,P'}=\int dY dY' dz dz' \bar \Phi
e^{i\frac{P+P'}{2}\cdot (Y-Y')+i\frac{k+k'}{2} \cdot (z-z')} e^{-i
\Delta \cdot (\frac{ \bar Y}{2} -\frac{ \bar b}{2})},\ea where $\bar
\Phi=<Y',s'| \psi_{\rho}(z') \bar \psi_{\sigma}(z)|Y,s>$,
and $$ \bar Y=(Y+ Y')/2; \:\:  \bar b= (z+z')/2.$$

A fundamental difference between ordinary parton distributions, and
the $GPD$, lies in the dependence on $\DD$.  We may interpret $\DD$ as
probing the dependence of the quark 2-point correlation on $\bar b
-\bar Y$, namely the {\it offset of the quark $CM$ coordinate relative
to the hadron $CM$ coordinate}.  The transverse component of $\bar b$
is the average {\it impact parameter} of the two quark fields
(explaining symbol ``$b$''.)  This concept cannot be formulated with
ordinary (diagonal) parton distributions evaluated at $\DD =0$.  In
ordinary parton formalism, the quark 2-point correlation may only
depend on the {\it difference between the locations of the two
fields.} This follows immediately from translational symmetry; as
$\Delta \ra 0$, translational invariance states that the dependence on
$\bar b-\bar Y$ cannot be conceived.  No wonder the $GPD$'s pose such
a puzzle when starting from the usual parton distribution.

The past few years have seen an explosion of interest in GPD's and an
overkill in short-distance expansions.  The meaning of these
expansions is that large virtual momenta $Q$ select small spatial
separations, namely the region of $z-z' \ra O(1/Q)$.  Here $Q$ is a
large scale such as a virtual photon's momentum serving as a pointlike
probe of the GPD. The short distance expansions themselves contain no
information about the target, and the mild $Q^{2}$ dependence of
logarithmic scaling violations is understood.  This is old physics; we
take as established that the quark operators in a large $Q$ reaction
can be viewed as nearly local and ``touching one another''.  This
concept is gauge-invariant.  When renormalizing the operators as a
function of $Q^{2}$, we may expect mixing to be diagonalized in $\bar
b$ space, because physics is {\it local}.  Meanwhile the {\it spatial
location } of the event inside the proton is absolutely not addressed
by the $Q^{2}$ dependence, should not be expanded via the operator
product expansion, and to reiterate, cannot be conceived or observed
with forward matrix elements.

{ \it New Partonic Amplitudes at Each Impact Parameter: }The $\DD^{+}$
dependence of reactions, usually denoted ``skewness'' in virtual
Compton parlance, is observable in particular frames; it is conjugate
to $(\bar z-v \bar t)$, the time dependence induced by a pulse of
moving fields in the Lorentz-boosted pancake.  Unfortunately the
$\DD^{+}$ dependence describes a vicinity in a pancake which is not
resolved better than the thickness of the pancake.  Localization in
the longitudinal direction is difficult.  Our view of the reactions
focuses on the $\bar b_{T} $ dependence.  Measurement of the $\DD_{T}$
dependence of amplitudes\cite{DGPR}, by Fourier transform, can be
inverted to find the spatial $\bar b_{T} $ location of the partons.
The transverse structure is directly {\it observable} when amplitudes
are measured by interference.  The transverse structure is remarkably
decoupled from the longitudinal smearing, Thus the region of
$0<\DD_{T}<GeV$, in which the handbag approximation is
consistent\cite{DGPR}, encodes the localization of various types,
flavors, momenta, transverse location and spin of partons, informing
us of these new quantities at each transverse locale.  The average
over the transverse plane of the imaginary parts of these partonic
amplitudes is the usual parton distributions at $\DD_{T}=0$.
Attention to the $\vec \DD_{T}$ dependence of reactions with GPDs at
large $Q^{2}$ can then reveal and localize the orbital angular
momentum of the quarks.

\section{The $F_{2}$ Form Factor and Orbital Angular Momentum}

Experiments on the proton's Pauli electromagnetic form factor $F_{2}$,
more than illustrate these ideas.  We have developed~\cite{QuebecF2} a
new approach to $F_{2}$, keeping in mind that the form factors have a
restrictive feature of $Q^{\mu}=\DD^{\mu}$.  This means that the
spatial resolution via form factors is never better than the region of
the object being measured, a limitation easily taken into account.

\begin{figure}
\epsfxsize=3in\epsfysize=2.4in

\epsfbox{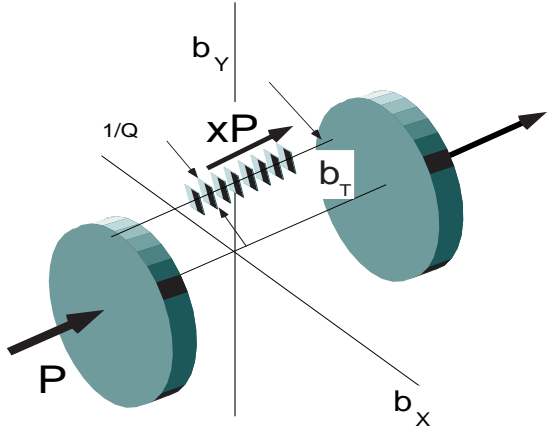}

\caption{Cartoon of the natural interpretation of $GPD$.  The fast
struck quark (stack of plane waves) is located at the transverse
offset position $\bar b_{T}$, while being spatially localized to within
order $1/Q$ in the transverse directions.  The quark's longitudinal
coordinate is not so well localized, but is spread along the
light-cone in the direction conjugate to $xP$.}

\end{figure}

We go to a (+,T,$-$) Lorentz frame in which the photon momentum
$Q^{\mu}=(0,\vec \DD_{T},0)$.  The initial and final proton momenta
are $P^{\mu}( P'^{\mu})= \\ (P,\mp \DD_{T}/2, \frac{m^{2}
+\DD_{T}^{2}/4}{2P}) $.  Physically $F_{2}$ represents a chirality
(helicity) flip amplitude $<-|J|+>$ where $|\pm> =|x> \pm i |y>$ are
chirality eigenstates.  Letting $x, y$ be two transverse orientations
of the chirality (``transchirality''), nearly equal to transverse spin
in the high energy limit, then $F_{2 } \sim <x|J|x> -<y | J|y> .$
Thus $F_{2}$ distinguishes the two possible transverse spin
orientations relative to the scattering plane $\vec \DD_{T}$.

We write the form factor as $$<P', s'| J^{\mu}| P,s>= \int
dk^{-}dxP^{+} d^{2}\bar b_{T} e^{i \DD_{T}\cdot \bar b_{T} /x} \;\bar
u_{i}(p',s')V^{\mu}_{ij}u_{j}(p, s) .  $$ We make the ansatz \ba
V_{ij}^{\mu} = ({\cal A} \gamma^{\mu} + {\cal B} \sigma^{\mu \nu}\bar
z_{\nu}+{\cal C} \sigma^{\lambda \nu}Q_{\nu}+{\cal D} \frac{(
P+P')^{\mu}}{m} \omega\gamma_{5} )_{ij}+\ldots.  \label{ansatz} \ea
The coefficients ${\cal A}\ldots{\cal D}$ are scalar functions of
$\DD^{2} , \bar b^{2}, \DD \cdot \bar b$, which under dimensional
scaling behave like $Q^{2}, \, 1/Q^{2},\, 1$, respectively.  At large
$Q^{2}$ we impose a factorization scheme forcing all diagrams into the
handbag: this unconventional step\cite{Brodsky89}  avoids the 
assumption that all
processes are necessarily factored into wave functions times hard
scattering.  Perhaps that assumption can be justified later; if so,
one relates the single unit of orbital angular momentum of the {\it
localized pair} of quarks to the interference of one and zero units in
wave functions, as we assumed elsewhere.~\cite{QuebecF2} We avoid
asserting that {\it all constituents of the proton} are localized at
short distances relative to one another due to any asymptotic
limit; our argument only requires the far more general property
that $\bar b_{T}$ of {\it one pair of operators} scales like $1/Q$.

We postponed defining the operator $\omega_{ij}=\Lambda
\epsilon_{\a \b \r \s} \sigma_{ij} ^{\a \b}\bar b^{\r}P^{\s}/m $ in
Eq. [\ref{ansatz}].  This operator selects orbital SO(2) harmonics $\bar
b_{x}\pm i \bar b_{y}= |\bar b_{T}|e^{\pm i \phi}$ which are good
under Lorentz boosts.  The operator also contains the non-relativistic
scalar $\bar b_{T} \times \vec p$ contracted with the Lorentz
generator $\sigma^{\a \b}$.  Spinors are $(\pm 1/2) $ eigenstates of
the Pauli-Lubanski operator $W^{\mu}s_{\mu}\gamma_{5} =\epsilon^{\mu \a
\b \r } \sigma_{\a \b} P_{\r} \,s_{\mu}\gamma_{5}/m $.  Separate the
component of $\bar b$ parallel to $s$ for this reason, writing $$\bar
b ^{\mu} = \bar b \cdot s \, s^{\mu} /(s^{2}) + \bar b^\mu_{r}.$$ Set
aside $\bar b_{r} $ orthogonal to $s$ for now; this makes $\omega
\gamma_{5} = \Lambda s_{T}\cdot \bar b_{T}$ on transversely polarized
states, and the $\cal D$ term is recognized as directly probing $\bar
b_{T} \times \vec p \cdot \vec s_{T}$. Note these coordinates apply
to the {\it pair} of operators and are gauge invariant.

In the limit of large enough $Q^{2}$ perturbation theory for three
quarks scattering yields a power series expansion of the form factors.
We apply that information to the handbag description to find bounds on
${\cal A}\ldots{\cal D}$.  In particular, the helicity-flip amplitudes
are power-suppressed compared to the helicity conserving ones, up to
Sudakov-corrections.  On this basis ${\cal D}$ and ${\cal A}$ are
comparable, although the numerical values depend on the relative
strength of the orbital $m=\pm 1$ terms compared to the $m=0$ terms.
The reasonable and difficult-to-avoid asymptotic scaling rule $\bar
b_{T} \ra Q_{T} / Q^{2}$ then gives $$F_{2} /F_{1}\sim \frac{
Q_{T}\cdot s_{T} }{ Q^{2} } .  $$ Of course $s_{T}$ appears in the
amplitude using the Pauli-Lubanski trick with the frames indicated,
and a direct calculation will give $F_{2} /F_{1}\sim 1/\sqrt{Q^{2}}$.

This result contradicts conventional wisdom, but it appears that
conventional wisdom overlooked orbital angular momentum.  The
distribution amplitude formalism of Brodsky and Lepage\cite{Brodsky89}
excludes orbital angular momentum in the first steps, leading to
hadronic helicity conservation as an exact dynamical selection rule,
and $F_{2}=0$.  Our calculation here parallels earlier
work,~\cite{QuebecF2} but is much more general.  After the meeting we
learned that Diehl, Feldman, Jacob and Kroll~\cite{Feldman} used GPDs
to study $F_{2}$.  The interpretation of interference between zero and
one unit of orbital angular momentum is explicit:see also Kroll's
report at this meeting\cite{KrollHere}.

Following an earlier CEBAF study~\cite{MJonesPRL} we corrected a math
error in our 1993 paper~\cite{Jainral93} introducing GPDs and relating
them to quark orbital angular momentum for $F_{2}$.  This led to a
prediction~\cite{QuebecF2} for $F_{2} /F_{1}\sim 1/\sqrt{Q^{2}}$.
Recent data from CEBAF extending to $Q^{2}= 6 \, GeV^{2}$ appears to
support this prediction.  The normalization of $QF_{2}/F_{1}$
indicates that the relative amplitudes for one and zero units of
orbital angular momentum are comparable.  Future studies accessible to
GPDs should show that the transversely polarized proton is not a small
round dot, but is flattened by the correlation of $s_{T}$ with the
scattering plane.  The power-law dependence predicted by quark
counting corresponds to a quadratic hole in the middle.  These
features should be more clearly visible when $Q^{2}$ and $\DD_{T}$ are
decoupled with DVCS experiments.~\cite{RalstonPire2001}: having
$Q^{2}>> \DD_{T} $ corresponds to examining the target on a
resolution small compared to the target size.

{\bf Acknowledgments:} Work supported by DOE grant number
DE-FGO2-98ER41079.

\section *{References}

\end{document}